\documentclass{article}
\usepackage[a4paper, margin=2.5cm]{geometry}
\usepackage{graphicx} %
\usepackage{siunitx}
\usepackage{layouts}
\usepackage{xspace}
\usepackage{amsmath}
\newcommand*{\siv}{SiV$^{-}$\xspace}
\newcommand*{\sivs}{SiV$^{-}$ centers\xspace}
\newcommand*{\darksiv}{SiV$^{2-}$\xspace}

\usepackage[backend=biber, style=phys, maxbibnames=5, minbibnames=1]{biblatex}
\DeclareFieldFormat[misc,online]{title}{#1}  %
\DeclareFieldFormat{eprint:arxiv}{%
  Preprint at \url{http://arXiv.org/#1}}

\DeclareFieldFormat[misc,online]{title}{#1}  %
\DeclareFieldFormat[article]{title}{#1}  %
\DeclareFieldFormat{journaltitle}{\mkbibemph{#1}}  %

\DeclareBibliographyDriver{misc}{%
  \usebibmacro{bibindex}%
  \usebibmacro{begentry}%
  \usebibmacro{author}%
  \setunit{\labelnamepunct}\newblock
  \printfield{title}%
  \newunit\newblock
  \usebibmacro{eprint}%
  \setunit{\addspace}%
  \printtext[parens]{\usebibmacro{date}}%
  \setunit{\adddot\addspace}%
  \usebibmacro{finentry}}

\title{Mitigating the Transition of SiV$^-$ in Diamond to an Optically Dark State}

\author
{Manuel Rieger,$^{1,2}$ Rubek Poudel,$^{1,2}$ Tobias Waldmann,$^{2,3}$ Lina M. Todenhagen,$^{1,2}$ \\ Stefan Kresta,$^{2,3}$ Nori N. Chavira Leal,$^{2,3}$ Viviana Villafañe,$^{2,3}$ \\ Martin S. Brandt,$^{1,2}$ Kai M{\" u}ller,$^{2,3}$ Jonathan J. Finley$^{1,2,*}$  \\
\\
\normalsize{$^{1}$Walter Schottky Institute and School of Natural Sciences,}\\ 
\normalsize{Technical University of Munich, 85748 Garching, Germany}\\
\normalsize{$^{2}$Munich Center for Quantum Science and Technology (MCQST), 80799 Munich, Germany}\\
\normalsize{$^{3}$Walter Schottky Institute and School of Computation, Information and Technology,}\\ 
\normalsize{Technical University of Munich, 85748 Garching, Germany}\\
}

\date{(Dated: December 5, 2025)}

\begin{document}

\maketitle

\section*{Abstract}

Negatively charged silicon vacancy centers in diamond (\siv) are promising for quantum photonic technologies. 
However, when subject to resonant optical excitation, they can inadvertently transfer into a zero-spin optically dark state.
We show that this unwanted change of charge state can be quickly reversed by the resonant laser itself in combination with static electric fields.
By defining interdigitated metallic contacts on the diamond surface, we increase the steady-state SiV$^-$ photoluminescence under resonant excitation by a factor $\ge3$ for most emitters, making it practically constant for certain individual emitters.
We electrically activate single \sivs near the positively biased electrode, which are entirely dark without applying local electric fields. 
Using time-resolved 3-color experiments, we show that the resonant laser not only excites the SiV$^-$, but also creates free holes that convert SiV$^{2-}$ to SiV$^-$ on a timescale of milliseconds.
Through analysis of several individual emitters, our results show that the degree of electrical charge state controllability differs between individual emitters, indicating that their local environment plays a key role. 
Our proposed electric-field-based stabilization scheme enhances deterministic charge state control in group-IV color centers and improves its understanding, offering a scalable path toward quantum applications such as entanglement generation and quantum key distribution.

\section{Introduction}

Quantum networks can potentially revolutionize distributed quantum technologies such as quantum communication, computation, and sensing~\cite{kimble2008quantum, wehner2018quantum}. A key requirement for building performant photonic quantum networks is the use of repeater stations. To realize them, long-lived qubit registers which process and store quantum information while counteracting fiber losses would be desirable~\cite{PhysRevX.10.021071}. Several physical platforms are under active investigation for these quantum nodes including trapped ions~\cite{sangouard2009quantum}, neutral atoms~\cite{covey2023quantum}, and color centers in diamond~\cite{wang2023field}. In particular, recent demonstrations of metropolitan scale quantum networks based on diamond spin systems confirm the suitability of such systems for solid-state quantum repeaters~\cite{knaut2024entanglement, stolk2024metropolitan}. 

Negatively charged silicon vacancy centers in diamond (\siv) exhibit outstanding photonic and spin properties, including a narrow zero phonon line (ZPL) at approximately \SI{1.68}{\electronvolt} ($\sim$\SI{737}{nm}), with up to 70\% of the emission occurring into the ZPL~\cite{muller2014optical}. For \sivs, the transmission losses in optical fibers can be reduced to \SI{0.2}{\decibel\per \kilo\meter} through bidirectional quantum frequency conversion to \SI{1550}{\nano\meter}~\cite{knaut2024entanglement}. Moreover, \sivs demonstrate exceptional spin properties, with electron spin coherence times reaching \SI{10}{ms} and nuclear spin coherence extending to several hundred milliseconds when employing dynamical decoupling techniques at sub-Kelvin temperatures~\cite{sukachev2017silicon}. \sivs consist of a silicon atom positioned interstitially within the diamond lattice, surrounded by two vacancies along the $\langle111\rangle$ crystallographic axis. Even though this inversion-symmetric structure protects against environmental electric noise~\cite{de2021investigation}, the charge state of the \sivs is influenced by multiple factors, including surface termination~\cite{Zhang2023PRL_SiV0-by-H-termination, grotz2012charge}, proximal dopants~\cite{Rose2018_SiV0_250msCoherence}, and illumination conditions, either through optical pulse sequences~\cite{TinVacancyCycle_Becher2022, Wood2023_SiV0_SiV2-_Meriles, Meriles_photoInducedChargeDynamics_2024}, or prolonged exposure to ultraviolet light~\cite{Pederson2024_aboveBandgapIllumination, Zuber2023_longIrradiation}. 

While charge stabilization via Fermi-level tuning is commonly used in other solid-state emitters such as quantum dots~\cite{Zhai2020_LDE_QDs,Hanschke2018_QDs_NPJ}, implementing this strategy in diamond poses additional challenges. This is primarily because diamond color centers typically lie deep within the large \SI{5.5}{\electronvolt} bandgap, where free carriers do not reach thermal equilibrium over experimentally relevant timescales. This is particularly true at low temperatures~\cite{Dhomkar2016_NV-long-time-chargeState-stability, Heremans2016_Review_DeepDefects}. 
Nevertheless, recent work has demonstrated that electric contacts placed on the diamond surface can actively modulate the SiV charge state by altering carrier capture and ionization rates~\cite{Dhomkar2018_SiV0-on-demand-doubtable,rieger2024fast}. Moreover, spatially resolved studies have revealed complex charge migration between diamond defects and nearby electrodes~\cite{Wood2023_SiV0_SiV2-_Meriles,Meriles_photoInducedChargeDynamics_2024}. Motivated by these findings, we recently investigated the underlying physical mechanisms driving the optoelectronic conversion between the optically dark \darksiv and bright \siv ensembles subject to \emph{off-resonant} optical excitation~\cite{rieger2024fast}. Our results demonstrated that the \siv charge state can be converted at \SI{}{\mega\hertz} rates by adjusting an external electric field. 

In this work, we present a hybrid optical and electrical stabilization protocol for \siv, engineered to stabilize the SiV$^-$ charge state under \emph{resonant} optical excitation. Using time-resolved experiments, we observe that \siv photoluminescence is quenched when subject to resonant excitation and confirm that this process can be reversed by applying a stabilizing external electric voltage from $20$ to \SI{120}{\volt} via a nearby electrode. To confirm whether the sole effect of the resonant laser is to excite the SiV$^-$, we add a third laser to the measurements, which is slightly detuned from resonance. We find that this near-resonant laser can also convert the dark state back to SiV$^-$, similar to the effect of green or blue lasers~\cite{TinVacancyCycle_Becher2022,SnV_BecherModel_Ikeda_IwasakiGroup,rieger2024fast}. We demonstrate that this applies to the resonant laser as well. However, only voltages from a certain range tip the balance towards stabilization of SiV$^-$ during resonant excitation, while the conversion to the dark state SiV$^{2-}$ dominates at the other voltages. 

Our findings are relevant for optical spin control protocols that rely on resonant excitation for spin initialization and readout, as well as for quasi-resonant Raman schemes that provide full control of the spin in the Bloch sphere. By mitigating the unwanted \siv transition to the dark \darksiv state with our proposed scheme, group-IV color centers can be more reliably scaled to larger quantum systems, facilitating multipartite entanglement generation and memory-enhanced quantum key distribution. In addition, our device architecture has minimal leakage currents, even at very high applied fields. This not only aids the charge state stability, but can also be used to tune the optical transition frequencies using the Stark effect while suppressing noise caused by fluctuating charges and photocurrent, as shown recently~\cite{rieger2025secondorderstarkshiftsexceeding}.

\section{Results}

\subsection{Baseline measurements on individual \siv without electrodes}

\begin{figure*}[h]
\centering
\includegraphics[width=\textwidth]{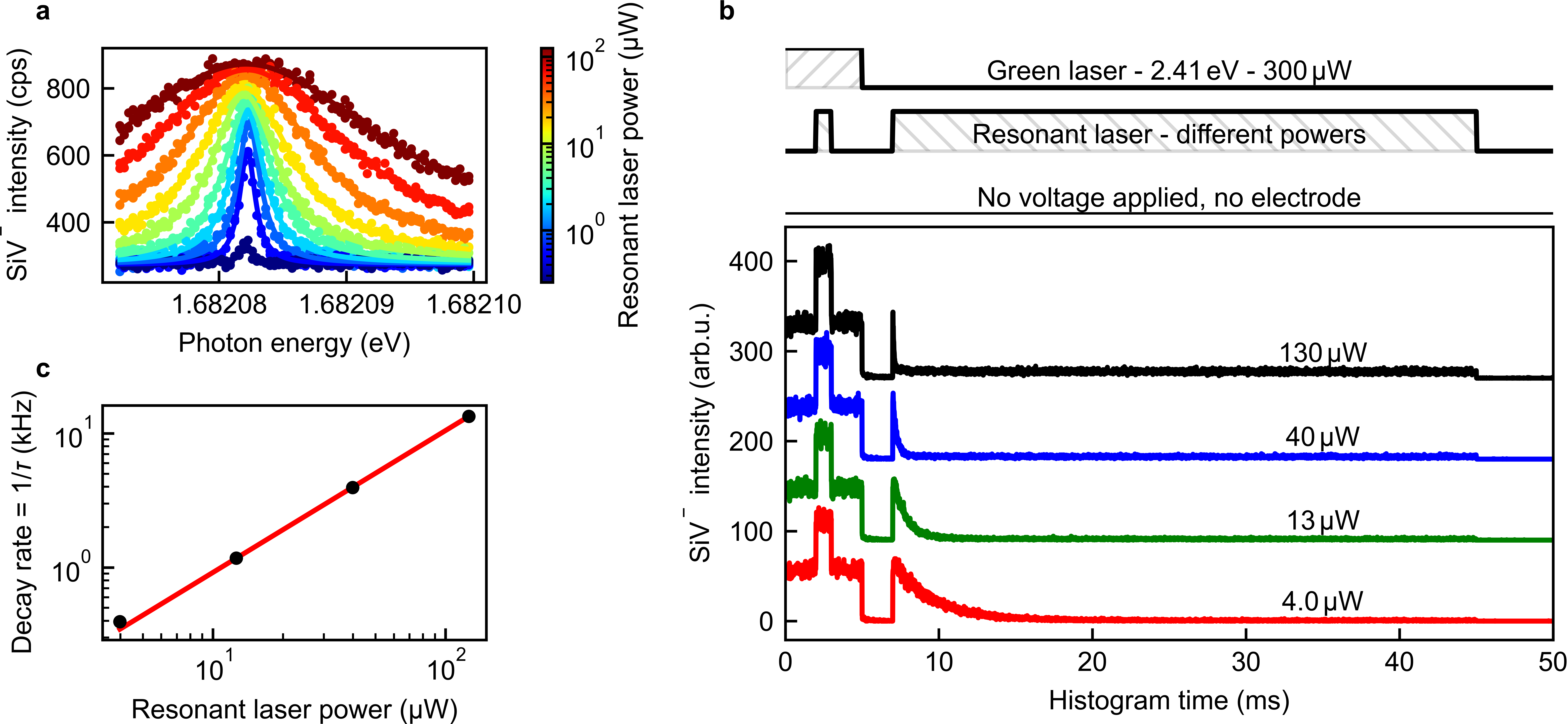}
\caption{\label{fig:Fig1}
\textbf{ A single \siv in pure bulk diamond under resonant excitation converts into a dark state.} 
\textbf{a} Photoluminescence excitation spectra as a function of resonant excitation power. The emitter PLE presents a Lorentzian line that saturates and broadens with increasing power. For stabilizing the emitter in the negative state, we use \SI{300}{\micro\watt} of a \SI{2.41}{\electronvolt}-photon energy laser. 
\textbf{b} Firstly, we apply an off-resonant stabilization laser for \SI{5}{\milli\second} with a short intermediate resonant probe, followed by a \SI{2}{\milli\second} pause. Then, we perform a \SI{38}{\milli\second} resonant readout. During the readout, the counts decay exponentially to almost zero intensity. 
\textbf{c} Fitted single exponential decay time for the decay from the bright \siv charge state to the dark state presented in panel~b. The decay time is linear in the resonant laser power. 
}
\end{figure*}

We begin our analysis by presenting photoluminescence excitation (PLE) spectroscopy of single emitters on an uncontacted diamond sample. For this, we simultaneously excite using a green laser with \SI{300}{\micro\watt} (\SI{0.7}{\micro\meter} $1/e^2$ Gaussian beam diameter) and a resonant laser with variable power (\SI{0.95}{\micro\meter} $1/e^2$ Gaussian beam diameter) and detect fluorescence from the excited state into the phonon side band, filtered in the spectral window between $1.55$ and \SI{1.65}{\electronvolt} photon energy (between $750$ and \SI{800}{\nano\meter}). Figure~\ref{fig:Fig1}a shows typical spectra possessing a peak centered at \SI{1.68}{\eV}~\cite{Neu2013_lowTempPLSiV,Hepp2014_PL_SiV,Evans2016_annealingProcedure}, as a function of the resonant laser power~\cite{Citron1977_PowerbroadeningExperimental}. For low resonant excitation powers, we observe a near-transform-limited linewidth of \SI{220(40)}{\mega\hertz}.

To characterize the \siv charge state lifetime under resonant excitation, we use time-correlated single-photon counting, averaging histogram data over many repetitions (integration time $\ge\SI{1}{\minute}$). A schematic of the experimental optical protocol used throughout this work is presented in the top panel of Figure~\ref{fig:Fig1}b. We begin by initializing the center into the \siv charge state using a \SI{5}{\milli\second}-long off-resonant green laser pulse (\SI{2.41}{\electronvolt}, \SI{300}{\micro\watt})~\cite{Goerlitz2020_SnV_phononSideband,rieger2024fast}. During this initialization pulse, a short resonant probe beam is applied to determine the \siv photoluminescence count rate subject to continuous optical stabilization. After the off-resonant green laser pulse, we introduce a \SI{2}{\ms} delay and subsequently switch on the resonant readout beam for \SI{38}{\milli\second}. The bottom panel of Figure~\ref{fig:Fig1}b presents a typical example of the obtained time traces. Initially, the detected photoluminescence count rate resembles that  observed under continuous off-resonant stabilization. However, we observe a fast decay of the fluorescence intensity from the bright \siv to the dark charge state over timescales $\le\SI{2.5}{\milli\second}$ (rate $\ge\SI{400}{\hertz}$). This timescale accelerates with resonant laser power. For all the measured time traces, the \siv photoluminescence decays to non-detectable levels after \SI{38}{\milli\second} of resonant illumination. We attribute this observation to charge conversion into the dark \darksiv state~\cite{TinVacancyCycle_Becher2022,rieger2024fast}. The \siv fluorescence is restored only after reapplying the off-resonant green laser. 

Following our analysis, we fit the time-resolved data with single exponential decay functions. Figure~\ref{fig:Fig1}c shows that the decay rate to the dark state increases linearly with the resonant excitation power, consistent with a single-photon ionization process of SiV$^-$ to SiV$^{2-}$. We note that selective excitation of a single ground-state orbital branch leads to orbital pumping, which occurs on a nanosecond timescale~\cite{Becker2016,Becker2018} and is therefore not expected to influence the much slower observed charge-state dynamics.

\subsection{Electrically enhanced luminescence of individual resonantly driven \sivs}

\begin{figure*}[!ht]
\centering
\includegraphics[width=\textwidth]{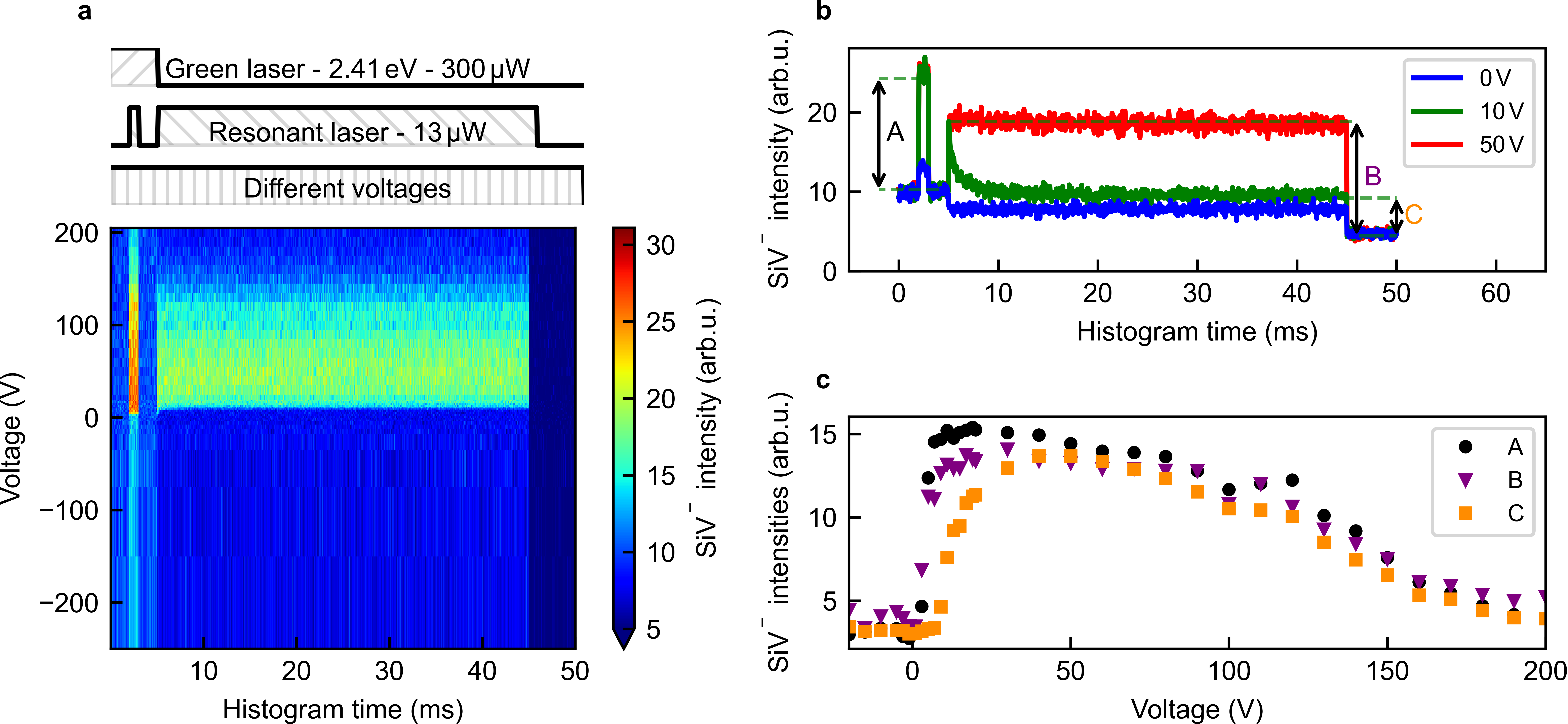} 
\caption{\label{fig:Fig2}
\textbf{ Charge state stabilization of a single SiV into its negative state. }
\textbf{a} We use the same optical pulse sequence as in Fig.~\ref{fig:Fig1} and, as indicated in the top panel, additionally apply an electric voltage. The resonant laser is tuned to the photoluminescence intensity maximum, measured at \SI{20}{\volt}. The emitter's intensity is negligible for negative voltages and seems highest between $0$ and \SI{+150}{\volt}.
\textbf{b} The \siv photoluminescence intensity decays within less than \SI{1}{\milli\second} at \SI{0}{\volt} while it disappears in approximately \SI{5}{\milli\second} at \SI{10}{\volt} and is almost perfectly stable at \SI{50}{\volt}. Double arrows indicate the intensity differences shown in the following panel for the example of \SI{10}{\volt}.
\textbf{c} The black circles indicate the \siv photoluminesccence countrate difference between combined resonant/off-resonant and purely resonant excitation. Purple triangles and orange squares are the background-subtracted countrates during the beginning and end of the resonant excitation pulse, respectively. The difference between the initial (purple) and the final count rate (orange) is almost zero above \SI{40}{\volt}, which indicates that the charge state is stable at these voltages. All intensities decrease at the highest voltages.
}
\end{figure*}

\begin{figure*}[!htb]
\centering
\includegraphics[width=\textwidth]{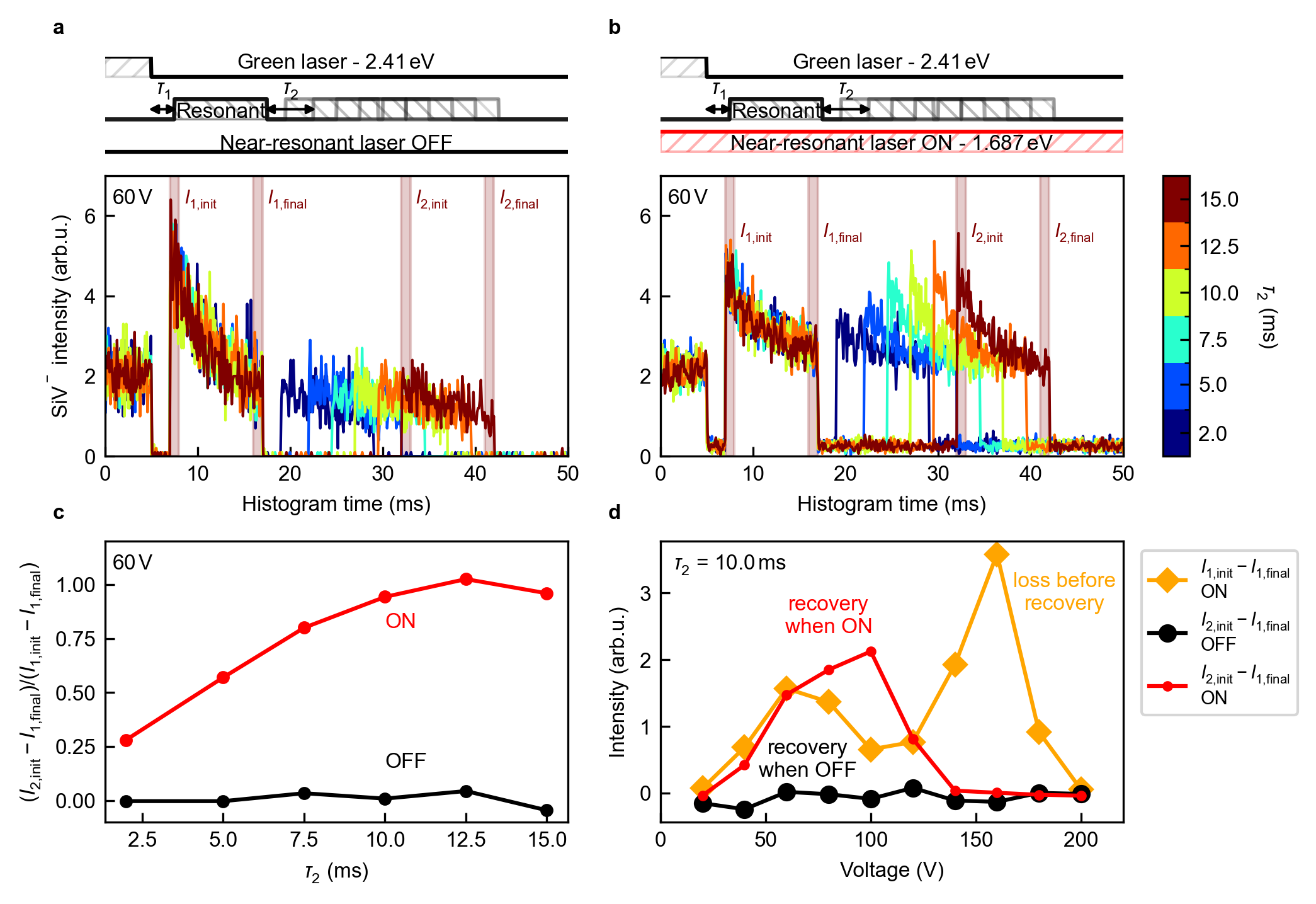}
\caption{\label{fig:Fig3} 
\textbf{ Time-resolved SiV$^-$ photoluminescence and charge state recovery of a single \siv induced by a near-resonant laser. }
A green laser pulse initializes the charge state into SiV$^-$. After a delay $\tau_1 = \SI{2}{\milli\second}$, two resonant pulses separated by variable delay $\tau_2$ excite the SiV$^-$. 
\textbf{a} Without the near-resonant laser, the SiV$^-$ photoluminescence decreases during resonant excitation and remains unchanged after the delay $\tau_2$. Red-shaded regions indicate \SI{1}{\milli\second} integration windows used to calculate intensities $I_{\text{1,init}}$, $I_{\text{1,final}}$, $I_{\text{2,init}}$, and $I_{\text{2,final}}$.
\textbf{b} With the near-resonant laser, the intensity increases progressively with increasing $\tau_2$, indicating charge state recovery to the SiV$^-$ charge state. While the near-resonant laser exhibits minimal direct excitation of the SiV$^-$ ground state (evidenced by low photoluminescence during $\tau_2$), it likely excites other defects or creates free carriers that facilitate charge state conversion back to SiV$^-$.
\textbf{c} Normalized recovery during $\tau_2$, calculated as the intensity recovered relative to the intensity lost during the first resonant pulse. Without the near-resonant laser, the normalized recovery remains near zero, but approaches unity within \SI{10}{\milli\second} of exposure to the near-resonant laser.
\textbf{d} Intensity dynamics at $\tau_2 = \SI{10}{\milli\second}$ for varying bias voltages. Orange diamonds show intensity loss during the first resonant pulse, large black circles show recovery without the near-resonant laser (negligible), and small red circles show recovery with the near-resonant laser. The near-resonant laser restores most of the lost intensity, with complete recovery achieved between \SI{20}{} and \SI{120}{\volt}.
}
\end{figure*}

We now continue to show that the application of an electric field effectively stabilizes the \siv charge state even under strong resonant optical pumping. For this, interdigitated titanium/gold electrodes are defined on the surface of the diamond, forming Schottky junctions. This structure allows high electric fields to be applied while maintaining low dark currents. 

We measure the highest current when we focus the green off-resonant laser on the edge of the positively biased electrode~\cite{rieger2024fast}. For all the measurements presented in this work, the electrode separation was \SI{7.6}{\micro\meter}, while the \siv position was typically \SI{1.9}{\micro\meter} ($=1/4$ of the electrode separation) away from the electrode edge. 
This spatial configuration ensures that the static electric field induced by the space charge region at the electrode/diamond interface remains small compared to the externally applied electric fields, which can be 20 times larger~\cite{rieger2025secondorderstarkshiftsexceeding}.
The optical pulse sequence employed is the same as that used in the previous section. However, the applied DC voltage creates a static applied electric field throughout the measurement.

Figure~\ref{fig:Fig2}a compares time-resolved photoluminescence histograms recorded as a function of the applied voltage $V$. Detectable photoluminescence emission is observed for $V<\SI{0}{\volt}$ or $>\SI{150}{\volt}$, indicating a progressive transition away from the bright \siv to the dark \darksiv state. This is consistent with the behavior described in Ref.~\cite{rieger2024fast}. For intermediate voltages, ranging from approximately $0$ to \SI{150}{\volt}, \siv photoluminescence is observed. 

Figure~\ref{fig:Fig2}b presents representative temporal traces recorded at 0, 10 and \SI{50}{\volt}. Here, we observe that the photoluminescence count rate reduces to the background level within the \SI{38}{\milli\second} resonant readout when the bias is set to \SI{0}{\volt}, consistent with the observations made in the previous section. At \SI{10}{\volt}, it is initially higher and subsequently exhibits a noticeable decay within a few milliseconds. Finally, when applying \SI{50}{\volt}, the photoluminescence count rate decreases by less than \SI{10}{\%} over the resonant \SI{38}{\milli\second}-long pulse, and is, therefore, essentially stable. We emphasize that even when we repeat the measurement with a ten times higher resonant optical power of \SI{130}{\micro\watt} as shown in Supplementary Fig.~4, %
the \siv charge state remains stable. In conclusion, these findings highlight that the local electric field induced by the gate voltage can stabilize the \siv charge state under resonant optical pumping. Thereby, our protocol ensures that (i) the \siv charge-state can be effectively initialized using off-resonant optical excitation, and (ii) the charge stabilization is persistent over at least tens of milliseconds.

Figure~\ref{fig:Fig2}c compares the \siv count rate during off-resonant optical stabilization with the initial and final count rates during the resonant pulse as a function of the applied bias voltage. Interestingly, the initialization appears to be slightly more efficient at \SI{20}{\volt}, suggesting that the optimal conditions for initialization and stabilization may not be identical. At \SI{0}{\volt} and negative voltages, the emitter is effectively dark, even at the beginning of the resonant readout, and the difference between the initial and final count rates is maximal up to \SI{20}{\volt}. These observations indicate that the emitter is activated, even when it would otherwise remain optically inactive in the absence of an applied bias, and that the electric field efficiently stabilizes its charge state.

To average the effects over many SiV centers, we performed ensemble measurements. In the Supplementary Note~3 %
we show two independent measurements with a silicon-implanted diamond that differs only in the 60x higher silicon implantation dose from the sample used for the rest of the measurements. Note that the higher implantation dose may increase the concentration of implantation-related crystal defects. The data show that applying positive voltages in the range of \SI{20}{} to \SI{50}{\volt} maximizes the \siv ensemble count rate after a green laser pulse as well as the steady-state count rate during resonant excitation. Additionally, they show that negative applied voltages quench the \siv ensemble count rate. Differences between the two measurements indicate that variations like electrode edge roughness or spatial variations in surface termination have an impact on the exact voltage-dependence of the \siv count rate. Overall, the data indicate that an applied voltage may improve charge initialization and charge stability of the majority of \siv centers.

We investigated six individual centers positioned at different spatial positions near the electrodes which are discussed in more detail in Supporting Note~2. %
Overall, these results reproduced the main finding that applying a voltage allows to stabilize the charge state in most of the studied \sivs. We identified two different voltage dependencies: Five emitters possess maximal brightness when applying voltages between \SI{30}{\volt} and \SI{60}{\volt} and can at least partially be stabilized by applying a voltage. Only one \siv emitter has maximal count rates at \SI{0}{\volt} and can not be stabilized. We tentatively ascribe the differences between emitters to the wide bandgap of diamond, and many diamond defects have energy levels that are so deep within the bandgap that they do not thermalize over the experimental timescales investigated~\cite{Dhomkar2016_NV-long-time-chargeState-stability, Heremans2016_Review_DeepDefects}. Therefore, the charge state of a center depends more on its local charge environment and illumination history than on the Fermi level position, which is only well defined in thermal equilibrium~\cite{LocalFermiLevel_Collins2002}. 

\subsection{The stabilizing effect of the resonant laser}

To investigate the origin of the charge state stabilization observed between $20$ and \SI{150}{\volt}, we designed an experiment to isolate the effects of the resonant excitation of SiV$^-$ and of the excitation of the surrounding charge environment. While the resonant laser can both directly excite the SiV$^-$ and may additionally modify the charge environment by exciting other defects to create free carriers, we isolate the latter effect using an additional near-resonant laser. This laser, blue-detuned by only \SI{4.6}{\milli\electronvolt} from the SiV$^-$ transition, has nearly the same photon energy but a significantly reduced SiV$^-$ excitation rate. We repeated the measurements with a red-detuned near-resonant laser which yielded qualitatively identical results.

We employ a pulse sequence consisting of a \SI{5}{\milli\second}-long green excitation pulse for initialization, followed by a fixed delay $\tau_1 = \SI{2}{\milli\second}$ and two \SI{10}{\milli\second} resonant pulses separated by a variable delay $\tau_2$. The SiV$^-$ photoluminescence is time-correlated with this pulse sequence to track charge state dynamics. The measurements are performed with a near-resonant laser, once turned off and once turned on at the same power as the resonant laser.

Figure~\ref{fig:Fig3}a shows the baseline behavior with the near-resonant laser off. The SiV$^-$ photoluminescence intensity decreases significantly during the first resonant pulse due to charge state conversion. Crucially, no recovery occurs during the delay $\tau_2$ and the intensity remains constant at its depleted level. 
Figure~\ref{fig:Fig3}b shows the same histogram when the near-resonant laser is on throughout the sequence. In stark contrast, we observe progressive intensity recovery during $\tau_2$. While the near-resonant laser induces minimal direct SiV$^-$ photoluminescence visible during the delay times, it excites other defects such as vacancies, divacancies and vacancy clusters~\cite{Davies1992,Pu2001_Divacancy_488nm}. This creates free charge carriers that alter the local charge environment and facilitates the conversion back to SiV$^-$, which requires free holes.

The recovery efficiency is quantified in Figure~\ref{fig:Fig3}c by normalizing the recovered intensity $(I_{2,\text{init}} - I_{1,\text{final}})$ to the initial loss $(I_{1,\text{init}} - I_{1,\text{final}})$. The respective intensities are indicated in Fig.~\ref{fig:Fig3}a-b. Without the near-resonant laser, recovery remains negligible across all delay times (the behavior changes for $>\SI{100}{\volt}$, see details below). With the near-resonant laser ON, recovery approaches unity for $\tau_2 \geq \SI{10}{\milli\second}$, demonstrating near-complete SiV$^-$ charge state restoration.

The voltage dependence in Figure~\ref{fig:Fig3}d reveals that intensity loss during resonant excitation peaks at \SI{60}{\volt} and \SI{160}{\volt}. Remarkably, the near-resonant laser recovers nearly all lost intensity between $20$ and \SI{120}{\volt} and even surpasses the green laser initialization between \SI{80}{\volt} and \SI{100}{\volt}. The voltage range in which the SiV$^-$ state is recovered by the near-resonant laser overlaps well with where we previously observed stabilized count rates (Figure~\ref{fig:Fig2}c). This confirms that the stabilization arises from the resonant laser's interaction with surrounding defects rather than purely electrical effects like Fermi level tuning or variations in the preparation of the charge environment by the applied voltage and off-resonant excitation. The applied voltage appears to enhance the net conversion rate from the dark charge state back to bright SiV$^-$, tipping the steady-state in favor of SiV$^-$.

Additional measurements discussed in the Supporting Note~1 %
show purely electrical charge state conversions. We find that below \SI{100}{\volt}, the charge states seem to be stable in the dark. However, voltages from \SI{100}{\volt} to \SI{120}{\volt} increase the SiV$^-$ population. Even higher voltages beyond \SI{130}{\volt} drastically reduce the \siv population without laser involvement. This aligns with \siv count rates dropping at such high voltages for the majority of the studied \siv centers.

\section{Discussion}

Our results demonstrate that electrical bias fields stabilize \siv centers by accelerating the recovery from the dark \darksiv charge state to \siv. For most of the studied centers, the applied electrical field improved not only the initialization with the green off-resonant laser pulse but also the \siv photoluminescence count rate under continuous resonant excitation. This provides a new degree of freedom for charge control, complementing all-optical methods while simultaneously enabling Stark tuning of the optical emission frequency. 
In addition, we demonstrate that near-resonant excitation, slightly detuned from the optical transitions, efficiently initializes the \siv charge state without requiring off-resonant lasers. This eliminates the need for additional laser sources and spectral filtering components, simplifying the optical architecture of quantum devices while reducing unwanted background fluorescence and phonon-mediated decoherence typically associated with off-resonant excitation.
We argue that the stabilization mechanism operates through voltage-driven hole photo currents that convert \darksiv to \siv. The excited state of the desired charge state \siv can convert to \darksiv, emitting a free hole which diffuses or drifts away. A laser charge-cycles other defects in the diamond, creating free holes which are attracted by \darksiv and convert them back to \siv. The applied voltage increases this photocurrent, especially near the positively biased electrode. Under continuous resonant excitation, a steady state then emerges between charge states SiV$^-$ and SiV$^{2-}$. This mechanism parallels the established model for off-resonant charge state stabilization~\cite{TinVacancyCycle_Becher2022,SnV_BecherModel_Ikeda_IwasakiGroup,rieger2024fast}. 
However, our data indicate that the resonant laser recovers \siv at a lower rate. Its reduced photon energy limits access to deeper defect states that the green laser can additionally ionize. This indicates that the most relevant defects for stabilizing SiV$^-$ have an optical ionization threshold between $1.68$ and \SI{2.41}{\electronvolt}. 
The optimal stabilization voltages ($20-\SI{150}{\volt}$ for most studied emitters) reflect a balance between two competing processes. Higher voltages drive higher (hole) currents, leading to larger hole capture rates of the dark SiV$^{2-}$ centers. This enhances the conversion rate to the bright SiV$^-$ charge state, increasing the SiV$^-$ population and its photoluminescence countrate in the steady state. 
However, the data show that above \SI{120}{\volt}, the \siv charge state converts to the dark state even in the absence of any laser, suggesting field-induced ionization or a tunneling process. 
The response to the applied voltage varies between the studied individual emitters. This reflects the individual defect environment around each SiV, which impacts the diffusion and drift of the free holes. In particular, SiV$^-$ centers appearing intrinsically charge-stable might only appear so because the recovery happens very fast. Specifically, when defects that readily provide free holes are so close to an SiV such that free holes are easily captured, then SiV$^-$ might be the prevalent charge state. This suggests that the best overall stabilization results will be reached by optimizing dopant concentrations and fine-tuning the individual emitters using an applied voltage.
Our findings align very well with recent density functional theory (DFT) calculations showing that both green and resonant excitation of SiV$^-$ can lead to SiV$^{2-}$ through excited-state ionization~\cite{claes2025chargecyclegroupiv}. This agrees very well with our finding that many SiV centers are not stable in the excited state of the \siv charge state, even under continuous green excitation and especially under resonant excitation. 

\section{Conclusion and outlook}

In summary, we showed that voltages applied to in-plane electrodes with micrometer separations can enhance initialization into the SiV$^-$ charge state and stabilize it for voltages from typically $20$ to $\SI{150}{\volt}$. This wide range of stabilizing voltages directly translates to a large potential Stark tuning range ideal for overcoming the inhomogeneous distribution of \siv emission frequencies. We demonstrate that the stabilizing effect is caused by faster re-conversion from the dark SiV$^{2-}$ to the bright SiV$^-$ state. This allowed us to activate and observe SiV$^-$ centers which would otherwise be hidden in their dark state. We could show that different emitters vary in the degree to which they respond to the stabilization voltage. This individuality is in agreement with the quite individual local charge environment around each SiV center, modulating the hole capture rate. 
Overall, the method presented here provides a promising avenue for mitigating the stochastically unstable charge state of SiV$^-$ centers.

Our experiments suggest future studies to investigate different electrode separations, SiV-to-electrode distances and other geometries like sharp electrode tips. Smaller distances and structures that concentrate the electric field locally could increase the local photocurrents, potentially enhancing the efficiency of the stabilizing effect of the lasers on the SiV$^-$ charge state. 
To further optimize the overall efficiency of the system, one could systematically optimize the initialization process using multiple laser colors and individual voltages for each initialization phase. Doping the diamond with defects that can emit free holes while charge cycling under illumination could improve the optical charge state stabilization efficiency.
Beyond single emitter control, our results suggest pathways toward increasing the efficiency when operating multiple \siv qubits in parallel. Individually addressable electrodes could compensate for emitter-specific variations and increase the yield of fully functional registers while maintaining global optical control. We believe that our results are a significant step in understanding the charge dynamics of SiV in diamond and for designing scalable SiV-based quantum technologies.

\section*{Materials and methods}

\subsection*{Sample preparation}\label{sec:sample}

To create silicon vacancy centers, we implanted silicon into high-purity, single-crystal, electronic grade diamond produced by Element Six. The diamond was grown using chemical vapor deposition (CVD) with a size of \SI{3}{\milli\meter} $\times$ \SI{3}{\milli\meter} $\times$  \SI{0.5}{\milli\meter}. The diamond contains nitrogen and boron with concentrations below $\SI{5}{ppb}$ and $\SI{0.5}{ppb}$, respectively~\cite{E6_handbook}. 

$^{28}$Si was implanted with \SI{132}{\kilo\electronvolt} by CuttingEdge Ions (expected implantation depth of \SI{100}{} to \SI{150}{\nano\meter}) with a dose of \SI{3e11}{ions/\centi\meter^{-2}} for the SiV$^-$ ensemble sample (only used for measurements in the Supplementary Note~3) %
and \SI{5e9}{ions/\centi\meter^{-2}} for the sample with individually resolvable SiV$^-$. After implantation, we annealed in several steps ($\SI{4}{\hour}$ ramp to \SI{400}{^\circ C}, $\SI{8}{\hour}$ at \SI{400}{^\circ C}, $\SI{12}{\hour}$ ramp up to \SI{800}{^\circ C}, $\SI{8}{\hour}$ at \SI{800}{^\circ C}, $\SI{12}{\hour}$ ramp to \SI{1100}{^\circ C}, $\SI{10}{\hour}$ at \SI{1100}{^\circ C}) under high vacuum ($<\SI{1e-7}{mbar}$)~\cite{Evans2016_annealingProcedure}. Throughout the process, all temperature increments were executed gradually to preserve a high vacuum within the annealing chamber, essential for avoiding surface graphitization of the diamond~\cite{ruf2019optically}.

We cleaned the sample in aqua regia ($\SI{1}{\hour}$ at \SI{90}{\celsius}) and concentrated sulphuric acid (\SI{1}{\hour} at \SI{225}{\celsius}), followed by exposure to an oxygen plasma which produces oxygen surface termination. Electrodes of \SI{7.6}{\micro\meter} distance were created by optical lithography and evaporation of Ti(\SI{5}{\nano\meter})/Au(\SI{80}{\nano\meter}) for the ensemble sample and Ti(\SI{20}{\nano\meter})/Au(\SI{200}{\nano\meter}) for the sample with individually resolvable SiV$^-$.

The interdigitated electrodes are each connected to \SI{200}{\micro\meter} times \SI{200}{\micro\meter} bond pads. The sample is glued to a well-thermalized copper plate and wire-bonded to a custom-made printed circuit board chip carrier. 

\subsection*{Optical and electrical measurements}

The sample was cooled to \SI{5}{\kelvin} using an attoDRY800 closed-cycle cryostat supplied by attocube systems AG.
We applied electrical voltage using a Keithley 2450 source-measure unit, which we also used to measure currents. We grounded the Keithley and half of the electrodes to the cold base plate of the cryostat in a star-shaped fashion.  This noticeably improved the spatial symmetry of \siv photoluminescence scans when comparing positive and negative applied voltages. Pulse patterns were created with a Pulse Streamer 8/2 from Swabian instruments. 

The optical measurements in this work were performed with a custom-built confocal microscope setup.  Off-resonant excitation employed Toptica Photonics SE iBeam smart diode lasers with either \SI{740}{\tera\hertz} (\SI{405}{\nano\meter}) or \SI{582}{\tera\hertz} (\SI{515}{\nano\meter}) optical frequency or a Laser quantum LLC torus laser with an optical frequency of \SI{564}{\tera\hertz} (\SI{532}{\nano\meter}).

For resonant excitation, we used a narrow-linewidth continuous wave laser (C-Wave GTR from Hübner photonics), wavelength-stabilized to a wavemeter with \SI{60}{\mega\hertz} absolute accuracy and \SI{2}{\mega\hertz} resolution (High finess WS7-60 IR-I). We stabilized its power using a software PID controller and analog modulation by a fast acusto-optic modulator (\SI{15}{\nano\second} rise time, driven with \SI{350}{\mega\hertz} in the first diffraction order). We sliced pulses from the laser using the same acousto-optic modulator with digital modulation.

To achieve spatial mode filtering, we coupled all excitation lasers in single-mode fibers.
We used a \SI{750}{\nano\meter} shortpass filter after combining both excitation lasers using a free-space dichroic mirror, to remove background from the excitation laser spectra.
We focused all lasers with an $f=\SI{2.87}{\milli\meter}$, $0.82$ NA, low-temperature apochromatic objective (LT-APO/VISIR/0.82 from attocube systems AG).

For detection, we used single-mode fiber-coupled avalanche photodiodes (APDs) with approximately \SI{700}{\hertz} dark counts by the company Excelitas Technologies Corp. or superconducting single photon detectors from Single Quantum with $<\SI{1}{\hertz}$ dark counts. The excitation laser was separated from the SiV$^-$ phonon sideband signal using three \SI{750}{\nano\meter} longpass filters and a \SI{550}{\nano\meter} longpass filter. A \SI{800}{\nano\meter} shortpass filter removed the diamond Raman signal of the \SI{737}{\nano\meter} resonant excitation laser which is at \SI{817}{\nano\meter}. 
Temporal correlations were recorded with the time-tagger QuTag from qutools or a TimeTagger X from Swabian instruments.

\subsection*{Use of large language models}

We acknowledge the use of ChatGPT by OpenAI and Claude by Anthropic for generating Python code snippets used for experimental control software and for the analysis and processing of experimental data for publication.

\printbibliography

\section*{Acknowledgements}

This work was supported by the BMFTR through projects SPINNING (13N16214) and epiNV (13N15702) and by the Bayerisches Staatsministerium für Wissenschaft und Kunst through project IQSense via the Munich Quantum Valley (MQV). %
We acknowledge support from the Deutsche Forschungsgemeinschaft (DFG, German Research Foundation) under projects PQET (INST 95/1654-1) and Germany’s Excellence Strategy – EXC-2111 – 390814868. %

We thank Mikołaj Metelski and Dr. Nathan Wilson for providing ScopeFoundryTURBO, the laboratory control software framework based on the public Scopefoundry python package that enabled these measurements, and for assisting us with its use. 
Additionally, we would like to thank HÜBNER Photonics, especially Dr. Sapida Akhundzada, for custom programming that helped us to automate measurements using the C-WAVE GTR laser system.

\section*{Author contributions}

M.R., V.V., M.S.B., K.M. and J.J.F. conceptualized the project. M.R., R.P., T.W. and N.N.C.L. performed the experiments. S.K. prepared the samples and fabricated the electrodes on the samples. M.R., R.P., T.W. and N.N.C.L. analyzed the data. All authors interpreted the data. V.V., M.S.B., K.M. and J.J.F. supervised the project. M.R. wrote the manuscript with input from all authors. 

\section*{Competing interests statement}

The authors declare no competing interests.

\end{document}